\documentclass[runningheads]{llncs}
\usepackage[T1]{fontenc}
\usepackage[utf8]{inputenc}
\usepackage{xcolor}
\usepackage{subfig}
\usepackage{longtable}
\usepackage{multirow}
\usepackage{subfig}
\usepackage{graphicx}
\usepackage[ruled]{algorithm2e}
\usepackage{algpseudocode}
\usepackage{amsmath}
\usepackage{amssymb}
\usepackage{wrapfig,booktabs}
\usepackage{float}
\usepackage{pifont}
\usepackage{url}
\usepackage{hyperref}
\usepackage{appendix}
\usepackage[subtle]{savetrees}

\begin{document}
\title{It Is Time To Steer: A Scalable Framework for Analysis-driven Attack Graph Generation}
\titlerunning{It Is Time To Steer: A Scalable Framework for Analysis-driven AG Generation}
%
\author{Alessandro Palma\inst{1}
\orcidID{0000-0002-9104-9904} 
\and
Marco Angelini\inst{2}
\orcidID{0000-0001-9051-6972}
}
\authorrunning{Palma et al.}
%
\institute{Sapienza University of Rome, Italy,
\email{palma@diag.uniroma1.it}
\and
Link Campus Univerisity, Italy,
\email{m.angelini@unilink.it}
}
\maketitle              
\begin{abstract}
Attack Graph (AG) represents the best-suited solution to support cyber risk assessment for multi-step attacks on computer networks, although their generation suffers from poor scalability due to their combinatorial complexity.
Current solutions propose to address the generation problem from the algorithmic perspective and postulate the analysis only after the generation is complete, thus implying too long waiting time before enabling analysis capabilities.
Additionally, they poorly capture the dynamic changes in the networks due to long generation times.
To mitigate these problems, this paper rethinks the classic AG analysis through a novel workflow in which the analyst can query the system anytime, thus enabling real-time analysis before the completion of the AG generation with quantifiable statistical significance.
Further, we introduce a mechanism to accelerate the generation by steering it with the analysis query.
To show the capabilities of the proposed framework, we perform an extensive quantitative validation and present a realistic case study on networks of unprecedented size. It demonstrates the advantages of our approach in terms of scalability and fitting to common attack path analyses.
\end{abstract}
\section{Introduction}
\label{sec:intro}

In today's digital world cyber attacks are becoming more complex to mitigate and have a higher impact on organizations' infrastructure.
In this scenario, it is crucial for any organization to promptly assess the potential cyber risks their networks are exposed to. 
This involves identifying the vulnerabilities in the network's systems and quantifying the risk of their potential exploitation~\cite{gonzalez-granadillo_dynamic_2018,9519490}.
Among the possible threat models, Attack Graph (AG)~\cite{kaynar_taxonomy_2016,zenitani_attack_2023} is a graph-based representation of the potential attack paths in a network used to analyze its cyber risk.
It is currently the primary threat model used by enterprises for threat analysis and risk management~\cite{stergiopoulos2022automatic}: among the others, Security Information and Event Management systems (e.g., IBM QRadar and LogRhythm) leverage AG for attack detection, 
network monitoring tools (e.g., Zeek and Suricata) integrate modules that generate AG to analyze malicious traffic, and
threat intelligence platforms (e.g., ThreatConnect and OpenCTI) use AG to map out threat actor tactics, techniques, and procedures~\cite{10.1007/978-3-031-17140-6_29}.

Although AGs are very expressive attack models, the above systems must face different problems.
The first one is the poorly scalable computation of the attack paths, even for networks of moderate size, as AGs may grow exponentially in the size of network hosts and vulnerabilities~\cite{kaynar_taxonomy_2016}.
Currently, much effort has been spent to find a trade-off between the accuracy of the model and computation time~\cite{zenitani_scalable_2022}.
Indeed, providing scalable approaches for attack path analysis is still an open problem~\cite{palma2023workflow,zenitani_attack_2023}.
%
A second emerging problem is that they postulate a classic analysis workflow in which the generation process comes first and the analysis comes next, aggravating the poor scalability of the generation process, as it slows down or completely stops the analysis in real scenarios~\cite{zenitani_attack_2023}.
%
A final problem concerns the alignment between the current network situation and the related AG model. While the classic workflow permits the attack path analysis on the complete set of attack steps, it also implies that any change in the network needs a re-computation of the whole attack graph to reflect such changes, exacerbating the scalability problem~\cite{gonzalez-granadillo_dynamic_2018}.

To mitigate these problems, we contribute a novel workflow for AG generation and attack path analysis.
It leverages the concept of \emph{statistical significance} to express the degree of trustfulness of a partial AG, where not all the attack steps are computed.
According to the foundational aspects of progressive data analysis~\cite{angelini2018review,fekete2016progressive}, it progressively supports attack path analysis already during the AG generation, allowing preventive analyses with quantitative quality-controlled statistically significant results.
To improve the accuracy of analysis performed on partial AGs, we design a \emph{steering approach} which automatically accelerates the AG generation based on the requested attack path analysis.
Given a path analysis query from an analyst, it models the queried attack path features, extracts from them the features of each single attack step, and uses them to prioritize the generation of attack paths that answer the analysis query.
We evaluate the speed-up of this process through a comprehensive scalability analysis performed on thousands of experiments, including different network scenarios and a real case study.
Summarizing, this paper contributes:
i) a novel framework for AG generation and attack path analysis based on progressive data analysis;
ii) a way to express statistical significance over partial AGs;
iii) a novel approach to steer the AG generation based on attack path analysis queries;
iv) a comprehensive validation through experimental evaluations for both statistically significant AG generation and the steering mechanism;
v) a case study demonstrating the possibility of managing large networks and conducting common attack path analyses.
The source code and all materials are available as open source repository\footnote{\url{https://github.com/XAIber-lab/ProgressiveAttackGraph}}.
\section{Preliminaries}
\label{sec:preliminaries}

\noindent {\bf Attack Graph Model.}
An Attack Graph (AG) is a graph-based representation of the possible paths an attacker can exploit to compromise an Information and Communications Technology (ICT) network.
It is based on two inputs~\cite{ingols_practical_2006,zenitani_attack_2023}:
the network \emph{reachability graph}, and
the network \emph{vulnerability inventory}.
The former is a directed graph where nodes are network hosts and edges are direct communication links between them.
%
A vulnerability inventory reports the list of network hosts with an associated set of vulnerabilities, typically obtained by vulnerability scanners and
security knowledge bases, as the National Vulnerability Database (NVD)\footnote{\url{https://nvd.nist.gov/}}.
%
With these inputs, AG modeling represents all the dependencies between hosts and vulnerabilities an attacker can exploit.
In this paper, we leverage the Topology Vulnerability Analysis (TVA) model~\cite{jajodia2009topological}, where the nodes are \emph{security conditions} and represent the attacker access privileges in a specific host, while the edges are \emph{exploit dependencies} and represent the possible movement of the attacker in case of a successful exploit.

\begin{definition}[Attack Graph]
    An Attack Graph $G=(V,E)$ is a directed multi-graph in which $V$ is the set of security condition nodes and $E$ is the set of labeled edges where an edge $e=(v_1,v_2,u) \in E$ indicates that the attacker can move from condition $v_1$ to condition $v_2$ by successfully exploiting vulnerability $u$.
\end{definition}

\noindent Once the AG model is defined, the next step is the \emph{Attack Graph Generation}, which is the computation of all the attack paths to analyze the attack surface.
They represent sequences of compromised devices and vulnerabilities exploited during the attack.

\begin{definition}[Attack Path]
    An Attack Path $AP=\langle v_1,u_{1,2},v_2,u_{2,3}, \cdots v_n \rangle$ is the ordered sequence of attacker states $v_1, \cdots, v_n$, interleaved by the sequence of vulnerabilities $u_{1,2}, \cdots, u_{n-1,n}$ which exploit allows an attacker to move between consecutive states.
\label{def:ap}
\end{definition}

\noindent {\bf Risk Model.}
When AG generation ends, cyber risks are associated with attack paths to analyze attack exposure or address specific requirements (e.g., damage to the network).
They are evaluated according to existing approaches~\cite{gonzalez-granadillo_dynamic_2018}, that consider CVSS metrics\footnote{\url{https://www.first.org/cvss/v3.1/specification-document}} to estimate the likelihood and impact of an attack path $AP$ and calculate the risk according to its standard definition: $risk(AP) = likelihood(AP) \cdot impact(AP)$.
The likelihood is calculated using the CVSS-3.1 exploitability metrics (Attack Vector, Attack Complexity, Privilege Required, and User Interaction).
The impact is determined by CVSS-3.1 impact metrics.
Likelihood, impact, and risk are in the range [0,1] (more details are available in the work by Gonzalez-Granadillo et al.~\cite{gonzalez-granadillo_dynamic_2018}).

\begin{definition}[Vulnerability and Attack Path Features]
Let $AP$ be an attack path and let $u_{1,2}, \cdots \in U_{AP}$ the sequence of its vulnerabilities.
We refer to \emph{vulnerability features} as the CVSS metrics of a single vulnerability $u_{i,i+1} \in U_{AP}$.
We refer to \emph{attack path features} as the metrics for evaluating the risk of attack path $AP$ (for the defined risk model they are likelihood and impact).
\end{definition}

\noindent Thus, an \emph{attack path analysis} is the answer to queries on attack path features, e.g., $Q=\{impact~\geq~0.9 \wedge likelihood<0.5\}$ represents the query for less probable ($<0.5$) but very dangerous ($\geq0.9$) attacks.
More formally:

\begin{definition}[Attack Path Query]
    An attack path Query $Q$ is a query that specifies a range of values for one or more attack path features, potentially in conjunction or disjunction, to express an information need over AG.
\end{definition}

\section{Overview of Our Approach}
\label{sec:framework}

\begin{figure*}[!ht]
    \centering
    \includegraphics[width=0.8\linewidth]{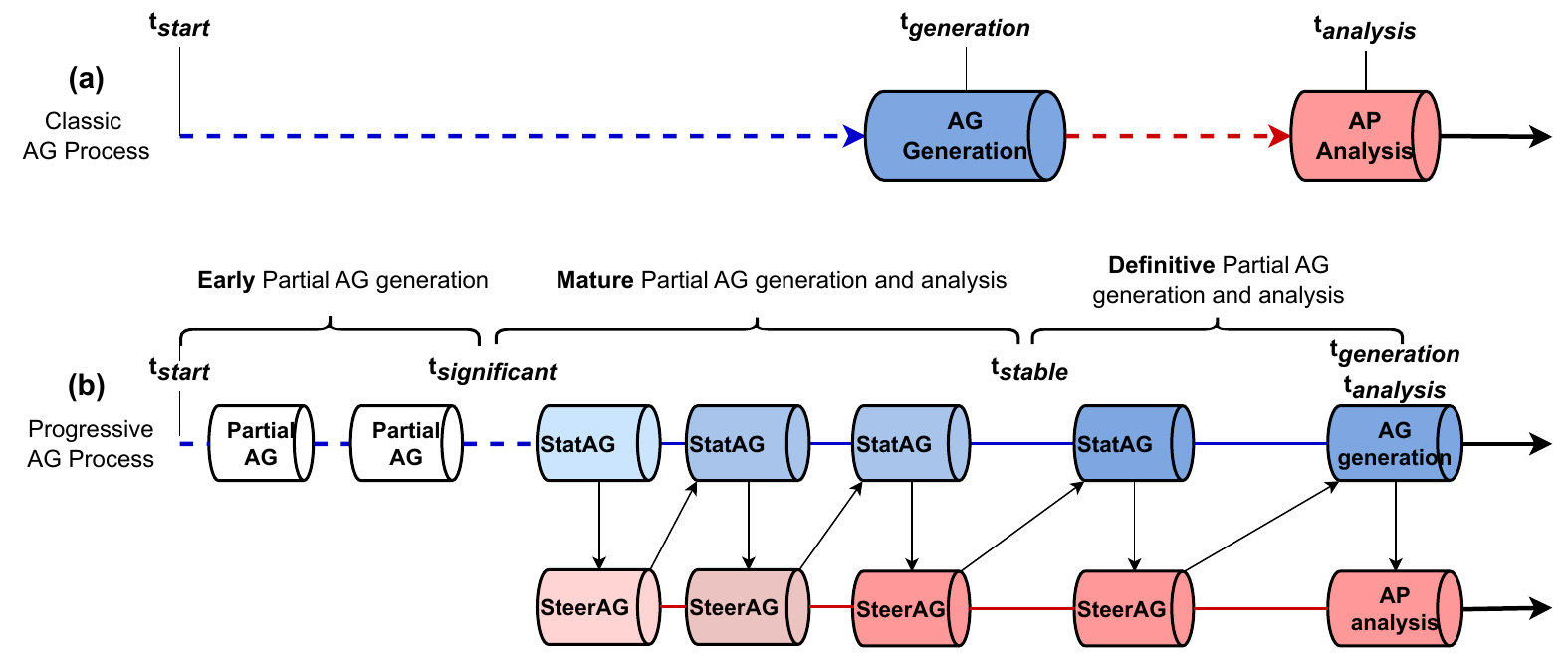}
    \caption{(a) Classic attack graph generation and analysis process and (b) the progressive one. Dashed lines indicate periods in which the analysis is stalled.}
    \label{fig:ag-workflow}
\end{figure*}

The classic process of attack path analysis~\cite{zenitani_attack_2023}, reported in Fig.~\ref{fig:ag-workflow}-a, comprises two milestones: the AG generation and the analysis of its attack paths.
The foundation of this framework is the representation of all the possible attack steps to provide a comprehensive view of the potential attacker's movements on the network.
The problem of enumerating the attack paths is combinatorial in the number of edges, determining one of the primary sources of their poor scalability.
As a consequence, the attack path analysis may never be performed because of this complexity, even for medium-sized networks, or the analyst should make assumptions in the generation, limiting the expressiveness and validity of the results.

In this paper, we propose a new framework for AG generation and analysis that leverages the fundamental concepts of \emph{progressive data analysis}~\cite{angelini2018review,fekete2016progressive}, that we report in Fig.~\ref{fig:ag-workflow}-b.
The main idea is to produce partial results during execution with increasing accuracy to generate intermediary outputs, potentially targeting interactive times.

In the case of AG generation and analysis, the framework we propose considers the computation of \emph{early partial} results that would give a coarse-grained approximation of the complete AG.
At a certain instant $t_{significant}$ (driven by the amount of processed data), the partial AG becomes \emph{statistically significant} (StatAG), in the sense that it is a good statistical approximation of the complete AG.
Statistical significance must consider the AG structure for its formulation.
From that moment on, attack path analysis generates \emph{mature partial} results, which reflect the outcome within an acceptable margin of error, quantifiable through statistical indicators.
This is the first advancement of the proposed framework, producing meaningful results to continuously explore the attack paths at interactive time.

As the progressive AG generation continues, there will be a certain instant $t_{stable}$ when the statistical significance reaches a threshold that guarantees \emph{definitive partial} results.
In these cases, the attack path analysis produces results that will no longer change substantially from the exact and, as such, supporting tasks like confirmatory analysis.
This final part runs until the complete AG is generated or paths are analyzed. 

This way of generating AGs allows the analysis of attack paths with preliminary data informed by quantifiable statistical indicators.
However, given the combinatorial size of AGs, it may still require a long time to reach the stability of definitive partial results.
For this reason, we contribute a \emph{steering mechanism} (SteerAG) to accelerate the convergence of the AG generation to the final outcome of any analysis executed during the generation~\cite{hografer_steering-by-example_2022}.
It consists of informing the next step of generation with the current attack path analysis.
For example, if the analysis query asks for the attack paths with the highest risk, then the steering mechanism prioritizes the generation of those high-risk paths.

To automatically steer the generation with the analysis query we cannot directly query the AG generator, as attack path features do not drive it. We leverage Machine Learning (ML) models to learn the vulnerability features from the structure of the attack paths and consequently prioritize the generation by selecting the vulnerability according to the learned features.
For this reason, the first phase of AG generation must be agnostic to the attack path analysis (white blocks in Fig.~\ref{fig:ag-workflow}-b) because it is used for collecting the initial balanced small set of attack paths that will be used to train the ML model and label attack paths as answers to the query or not.
Consequently, the ML model is trained with the collected labeled attack paths to predict the vulnerability features representative only of attack paths that answer the query positively.
%
This steering mechanism allows a faster convergence of the partial AG to the portion of the complete AG that answers the attack path analysis query.
In the next two sections, we describe the details of the statistically significant AG generation and the steering mechanism, along with their validation.
\section{StatAG: Statistically Significant Generation}
\label{sec:stat_ag}
\begin{wrapfigure}{r}{0.5\textwidth}
    \vspace{-20pt}
    \includegraphics[width=\linewidth]{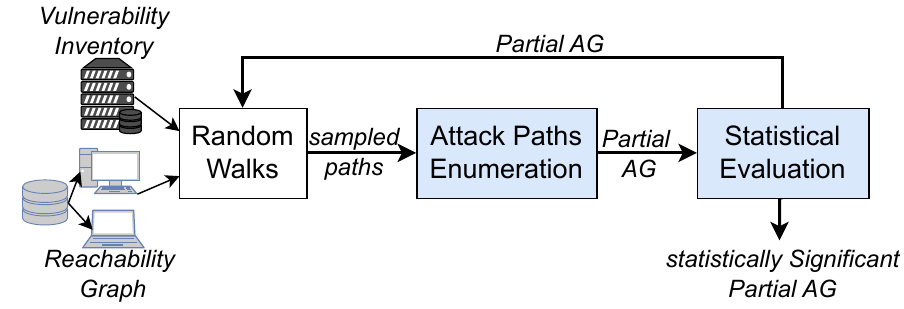}
    \caption{StatAG workflow.}
    \label{fig:wf-stats}
    \vspace{-20pt}
\end{wrapfigure}
This section describes how to achieve a progressively convergent statistically significant AG generation. 
It enables a new way of managing AGs with the main advantage of allowing the exploration and analysis of attack paths in partial AGs without waiting for the complete generation.
The core idea of this approach, named \emph{Statistically Significant Attack Graph Generation} (StatAG), is considering partial AGs being aware of statistical significance that becomes gradually higher until it reaches its maximum value corresponding to the complete AG (i.e., the exact result).
Fig.~\ref{fig:wf-stats} reports the workflow of StatAG.
%
%
The main activity during AG generation is the enumeration of its attack paths.
%
While existing approaches 
leverage common algorithms for graph traversal, i.e., BFS or DFS, 
we generate attack paths using random walks~\cite{li2015random} over the reachability graph.
A random walk is a path made of a succession of random steps in the graph.
The rationale for random walks is their capability of generating unbiased samples, contrary to the classic traversal algorithms, which depend on the starting node.
Thus, they facilitate a quicker convergence of the path features of the partial AG to the complete one~\cite{li2015random}.

The next step involves constructing the corresponding attack path.
Let us remember that the modeled AG is a multi-graph. Therefore, multiple edges (i.e., vulnerabilities) may exist between two nodes, although only one of the multi-edges can be part of an attack path, according to Definition~\ref{def:ap}.
When this happens, the construction of the attack path involves randomly selecting one vulnerability uniformly at random so that the total number of attack paths aligns with the number of sampled walks.
The collection of attack paths constructed from the sampled walks defines the partial AG.

\begin{definition}[Partial Attack Graph]
    Let $G=(V,E)$ be a complete Attack Graph, then a Partial Attack Graph $PAG_i=(V_i,E_i)$ is composed of a subset of nodes $V_i \subset V$ and edges $E_i \subset E$ retrieved after $i$ iterations of random walk sampling.
\end{definition}

\noindent At this point, the approach iterates toward the generation of other partial AGs that are progressively added to reconstruct the complete AG, eventually.
At the end of each iteration, we evaluate the statistical significance of the partial AG that quantifies the degree of its approximation to the complete AG. 
If it is acceptable, then the partial AG can be used for the initial exploration of the AG, along with indicators communicating its degree of uncertainty.
To measure the statistical significance, we need to define the \emph{null hypothesis} $H_0$, which claims that no relationship exists between two sets of data being analyzed~\cite{sproull2002handbook}.
In the context of the proposed approach, the two sets of data are the attack path features of the partial AG and the complete AG, and the statistical significance refers to the probability of rejecting the null hypothesis.
To evaluate this probability, we need to measure the probability $p$ of obtaining test results at least as extreme as the observed results (namely \emph{p-value}) and the probability $\alpha$ of rejecting the null hypothesis (namely \emph{significance level}).
According to these definitions, the partial AG is statistically significant when the $p \leq \alpha$, with the value of $\alpha$ commonly set to 0.05~\cite{sproull2002handbook}.

We evaluate the null hypothesis through the Kolmogorov-Smirnov (KS) statistical test, which compares two data distributions and quantifies the distance between them~\cite{massey1951kolmogorov}.
In particular, we evaluate the distance between the attack path feature distributions of the complete and partial AGs.
Let $\mathcal{D}_{AG}(x)$ and $\mathcal{D}_{PAG}(x)$ be the distribution of the attack path feature $x$ according to the complete and partial AG, respectively.
The KS distance of the two distributions is:
\begin{equation}
    KS(x) = sup_{x} \mid \mathcal{D}_{AG}(x) - \mathcal{D}_{PAG}(x) \mid,
    \label{eq:ks_distance}
\end{equation}
where $sup_{x}$ is the supremum function, 
which corresponds to the least upper bound of the distances between the two distributions of $x$.
According to this definition, we can define the statistical significance of partial AGs.

\begin{definition}[Statistical Significance for partial AG]
    Let $AG$ be the complete attack graph and let $PAG_i$ be the partial attack graph generated at the $i^{th}$ iteration.
    Let $H_0$: $KS(x) > T$ be the null hypothesis where $KS(x)$ is the KS distance between the distributions of the attack path feature $x$ for the complete and partial attack graph, and $T$ a predefined distance threshold.
    Then, $PAG_i$ is \emph{statistically significant} for the attack path feature $x$ if the $p \leq \alpha$, with $p$ the p-value of the $KS(x)$ and $\alpha$ the significance level.
    \label{def:statistical_significance}
\end{definition}

\noindent Let us note that this definition is appropriate only for a posteriori evaluation of the statistical significance because it requires the complete attack graph $AG$.
To estimate the statistical significance in a real-time application, we need a way to provide its measure during the progressive execution of the approach.
For this purpose, we introduce the concept of \emph{Attack Path Feature stability} (or simply stability) to quantify the variability of the partial results of an attack path feature gathered during the different iterations.
To measure the stability of an attack path feature $x$, we consider the KS distance between the cumulative distribution of $x$ until the $i^{th}$ iteration and the one at iteration $i+1$.
It indicates how much the new samples of iteration $i+1$ vary the already sampled distribution, with the rationale that a higher KS distance corresponds to more unstable results given the higher variability.

\begin{definition}[Attack Path Feature Stability]
    Let $PAG_{i}$ be the partial attack graph after $i$ iterations of the approach and $\mathcal{D}_{PAG_{i}}(x)$ the cumulative distribution of the attack path feature $x$ in $PAG_{i}$.
    The stability $\Delta$ of $x$ over $PAG_{i}$ is:
   \begin{equation}
       \Delta_{PAG_{i}}(x) = 1 - \mid \mathcal{D}_{PAG_{i-1}}(x) - \mathcal{D}_{PAG_{i}}(x) \mid
    \label{eq:stability}
   \end{equation}
\end{definition}
%
Given that the KS distances are defined in the range [0,1], we subtract 1 from the difference of the distances to express the rationale that a higher stability value corresponds to a more significant partial AG.

In the rest of this section, we validate the theoretical formulation of statistical significance and its correlation with the attack path feature stability.

\subsection{StatAG Validation}
To validate StatAG we present the experimental setting, 
then we analyze the convergence of vulnerability and attack path features, and stability distributions.

\noindent {\bf Experimental Setting.}\label{sec:setting_config}
We validate StatAG with an experimental setup designed for approaching the different factors that affect the AG scalability~\cite{zenitani_attack_2023}.
It consists of synthetic networks and vulnerability inventories, in which we vary the number of hosts, vulnerabilities, network topology, and composition of vulnerabilities per host. Looking at the former, we considered configurations with up to 20 hosts and 50 vulnerabilities per host, which is the biggest configuration possible to compute all the paths (complete AG, used as ground truth) in a reasonable amount of time (in the order of hundreds of hours)~\cite{sun_heuristic_2022}.
We then tested $3$ types of network topologies, including mesh, random, and power law.
We considered $5$ levels of heterogeneity of vulnerabilities, varying from 0\% (i.e., all vulnerabilities among the hosts are the same) to 100\% (i.e., all vulnerabilities among the hosts are different) to increase the variability of vulnerability features and $4$ vulnerability distributions (Uniform, Pareto, Binomial, and Poisson) to simulate realistic scenarios.
Thus, for each network, we evaluated 60 different configurations, which we varied in 100 statistical variations for a total of 6000 experiments.
We use the scikit-learn library~\cite{JMLR:v12:pedregosa11a} for the implementation and we run the experiments on a Linux server with Intel(R) Xeon(R) Gold 6248 CPU 2.50GHz and 256 GB memory.

\noindent {\bf Results.}
We validate StatAG by studying the convergence of the partial AGs to the complete AG.
For the sake of simplicity, in the rest of this section, we refer to the complete AG as \emph{ground truth} (GT) to indicate that it corresponds to all the paths generated at the end of the classic generation process.
\begin{figure*}[!ht]
    \centering
    \includegraphics[width=0.9\linewidth]{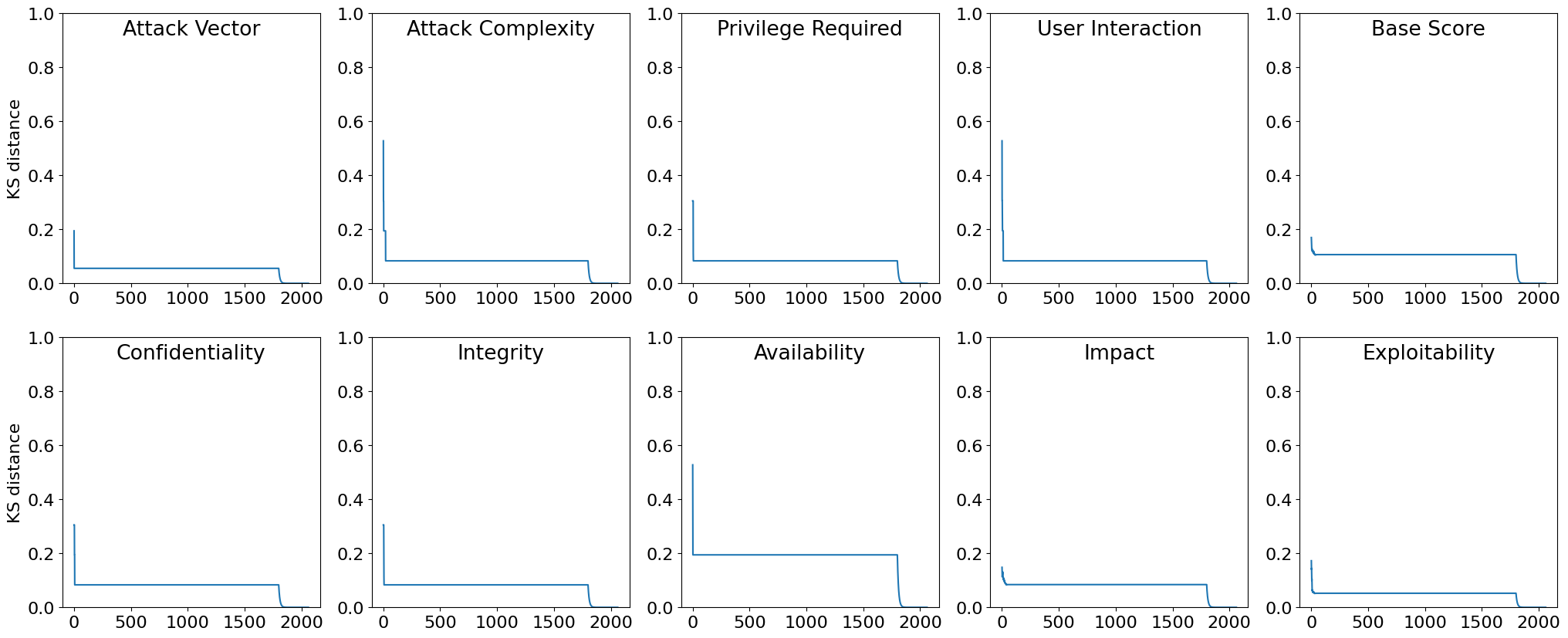}
    \caption{KS distance of partial AGs from the GT (vulnerability features).}
    \label{fig:base-stats}
\end{figure*}
The first validation step concerns the convergence of the vulnerability feature distributions (from CVSS) of the partial AGs to the GT.
Fig.~\ref{fig:base-stats} reports the median KS distances during the entire approach execution.
The convergence speed of the vulnerability feature distributions depends on the distance threshold $T$ (see Definition~\ref{def:statistical_significance}).
For example, considering a good approximation as $T=0.1$ (corresponding to just 10\% approximation of the GT), 9 out of 10 features are statistically significant after just 100 iterations, except for the availability feature that has a KS distance of 0.2 after 100 iterations.
In particular, we can notice three different trends in all the plots in Fig.~\ref{fig:base-stats}.
The first one coincides with early partial results, where the KS distance from the GT is too high to perform analysis with a reasonable approximation but suitable for initial exploration. This trend lasts about the first 100 iterations.
After 100 iterations, corresponding to 9\% of the complete execution, the trend is flat until the 1700-th iteration, corresponding to 77\% of the complete execution.
\begin{wrapfigure}{r}{0.5\textwidth}
    \vspace{-20pt}
    \includegraphics[width=\linewidth]{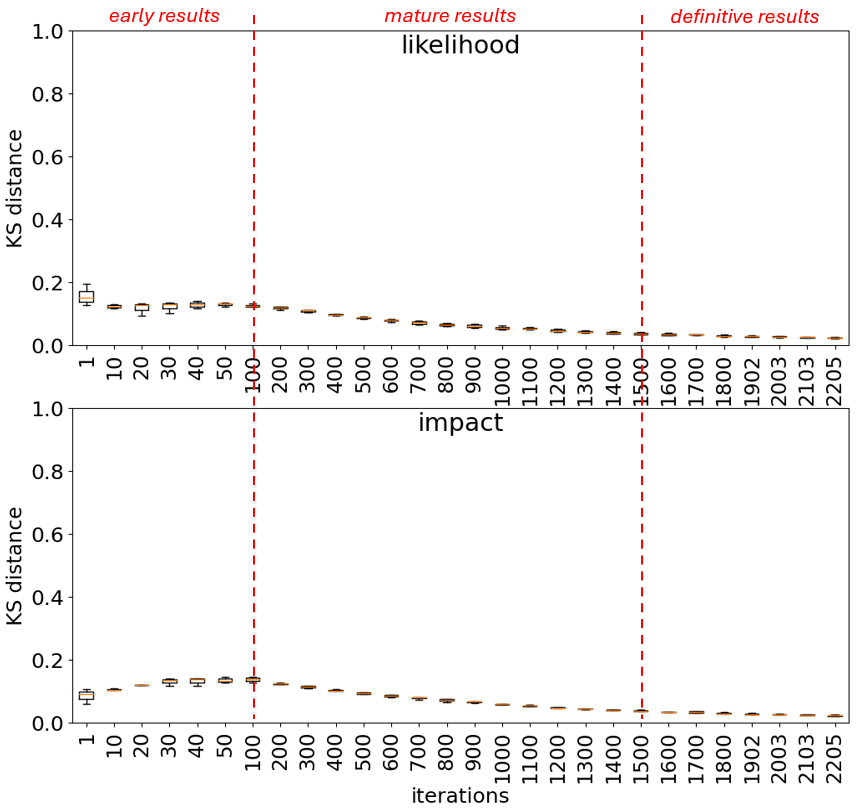}
    \caption{KS distance of partial AGs from GT (attack path features).}
    \label{fig:path-stats}
    \vspace{-25pt}
\end{wrapfigure}
This is the interval where mature partial results appear, that provide a 10\% approximation of the GT for most vulnerability features, with the worst performance reached by the availability with a 20\% approximation. 
Finally, in the last 23\% of the execution, the KS distance becomes very close to 0, corresponding to the definitive partial results.
Consequently, the vulnerability variability is captured in just 9\% of the total iterations (the moment in which mature partial results appear), identifying the iteration from which an analyst can start reasoning on the AG.
The convergence trend of the vulnerability features to the GT shows the capability of the progressive approach to represent 90\% of the vulnerability inventory after a few iterations.
It is a good result for the analysis of the vulnerability inventory, but we need further investigation that concerns the attack path analysis.

To this aim, the next validation step concerns the convergence trend of the attack path features of the partial AGs to the GT ones.
Fig.~\ref{fig:path-stats} reports the distribution trend of the KS distances between the partial AGs and the GT for the attack path features, i.e., likelihood and impact.
%
The trends show a gradual convergence for both likelihood and impact, with the largest distance equal to 0.2 in the first 100 iterations, where the variability of the data is also higher (i.e., the box size of the boxplots).
The KS distance distribution becomes less variable and with lower median values after about 100 iterations (9\% of the execution) and until 1500 iterations (68\% of the execution).
It is the case of mature partial results, that approximate the attack path features distribution from 15\% to 5\% accuracy (i.e., KS distances from 0.15 to 0.05).
After 1500 iterations and until the end, we can recognize definitive partial results with a KS distance very close to 0.
In summary, attack path features can be queried after a few iterations with a 20\% approximation, gradually decreasing, reaching a 5\% approximation after 1500 iterations, comparable to vulnerability features (Fig.~\ref{fig:base-stats}).

\begin{wrapfigure}{l}{0.5\textwidth}
    \vspace{-20pt}
    \includegraphics[width=\linewidth]{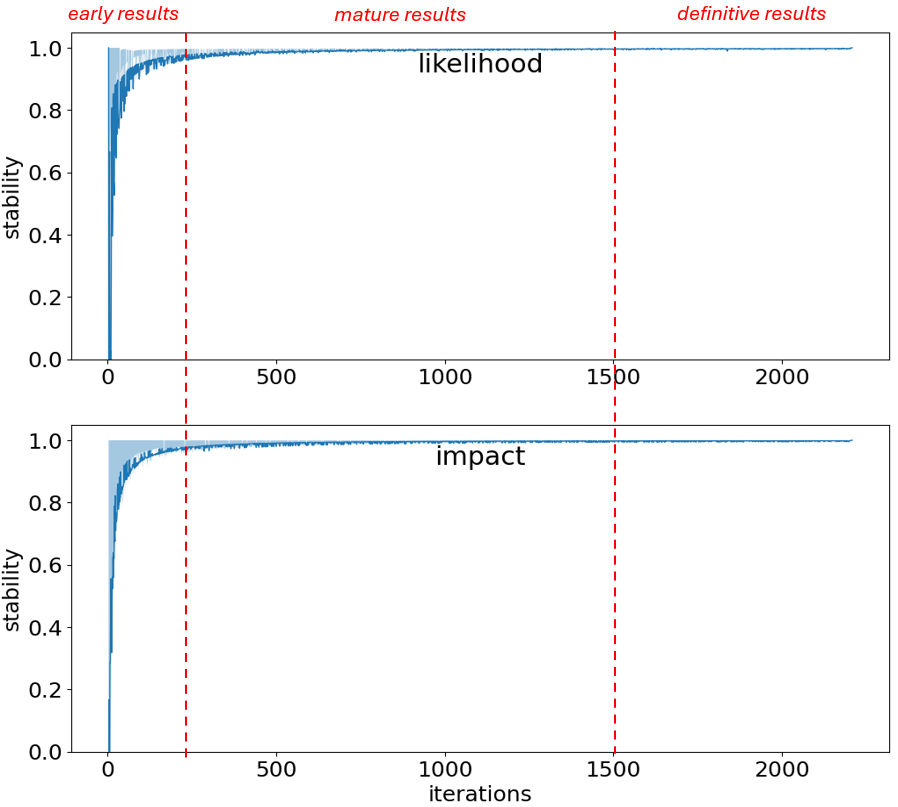}
    \caption{Stability for the path features.}
    \label{fig:stability}
    \vspace{-20pt}
\end{wrapfigure}

In the final validation step, we furnish evidence demonstrating that the attack path feature stability (Equation~\ref{eq:stability}) accurately reflects the trend in statistical significance so that it can be used as a real-time indicator of the convergence to the GT.
To this aim, Fig.~\ref{fig:stability} reports the stability trend for the performed experiments, where the trend of the median values is represented with full-color hue, while the alpha blended areas identify upper and lower quartile values to report the statistical variability.
The stability trend is coherent with the KS distance trend of Fig.~\ref{fig:path-stats}: during the first 100 iterations (9\% of the execution), the stability indicates early partial results as well as the KS distance with the GT, with stability values lower than 0.85.
Between iterations 100 and 1000 (46\% of the execution), the stability reports an intermittent trend between 0.85 and 0.95, indicating the mature partial results, where the AG is assisting its statistical significance. The same trend corresponds to the KS distance from the GT.
Finally, we can identify the definitive partial results in the last 54\% of iterations, where the stability is very close to 1 and, correspondingly, the KS distance with the GT is very close to 0.

In conclusion, this validation experimentally proves the correlation between stability and statistical significance in different configurations.
While statistical significance can be calculated only a posteriori, stability has the great advantage of being computed in real-time, thus informing the analyst at each iteration.
%
%
The StatAG validation showed the ability of the approach to generate the attack paths progressively, with three thresholds of significance (early, mature, and definitive) that inform about the convergence of the partial AG to the GT.
This progressive approach represents an innovation in the Attack Graph community, where traditional solutions typically focus on the complete generation of AGs. It allows continuously fed data from the AG at every iteration and is more efficient in computing a statistically representative version of the AG.
However, when it comes to analyzing the complete AG for attack path analysis, this approach still requires enumerating all paths.
To expedite convergence towards the subset of the complete AG that addresses a particular attack path query, the next section introduces a steering mechanism designed to guide both the generation and analysis processes.

\section{SteerAG: Steered Generation and Analysis}
\label{sec:steering_ag}
\begin{wrapfigure}{r}{0.4\textwidth}
    \vspace{-20pt}
    \includegraphics[width=\linewidth]{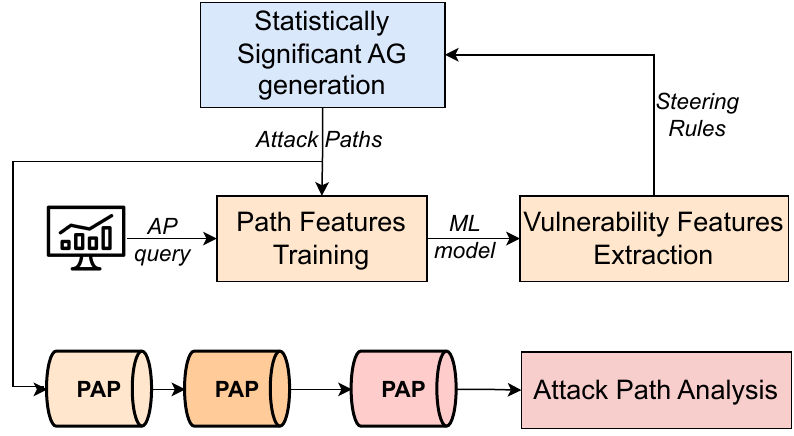}
    \caption{SteerAG workflow.}
    \label{fig:wf-steer}
    \vspace{-20pt}
\end{wrapfigure}
%
StatAG provides an approximation of the entire AG during each iteration, only partially solving the scalability issue, as exact results may still take a long time to obtain for some queries due to low convergence speed despite offering varying result qualities for different analyses.
For this reason, we contribute a further improvement of the framework, central in its capability to merge analysis and generation, and let the second be guided by the first, consisting of accelerating the attack path analysis for dynamic and real-time environments through the design of a steering mechanism, reported in Fig.~\ref{fig:wf-steer}.
We name this approach \emph{Steered Attack Graph Generation and Analysis} (SteerAG).
According to the framework (see Fig.~\ref{fig:ag-workflow}), the steering process is activated with the presence of partial attack paths (PAP in Fig.~\ref{fig:wf-steer}) generated using StatAG and an attack path query for the analysis, issued by a human user or automatic process.
%
%

The first step of the steering approach is the \emph{path features training}, which consists of two activities.
First, we label the attack paths coming from the StatAG as ``relevant'' when their attack path features correspond to the query requirements and ``not relevant'' otherwise. As this operation is linear in the number of attack paths generated up to the issuing of an analysis query in the worst case, it does not represent a costly operation.
When a sufficient number of attack paths are labeled, with this number depending on the ML model used (i.e., ten to fifty for decision trees, around 100 for neural networks), the second activity is the actual training of a binary ML-based classification model, where the attack paths are classified according to their assigned label. The classification is based on the vulnerability features of the paths and not on the characteristics of the whole path. In this way, we aim to discover the latent relations between all the relevant attack paths and their vulnerability characteristics based on the ones available. Extracting these relations allows for steering the AG generator by the discovered intervals of vulnerability features.

In our proposal, we use the decision tree classifier~\cite{kotsiantis2013decision} as the ML model for its properties that fit well with the steering approach as they:
(i) can be trained quickly and therefore are suitable for progressive data analysis~\cite{hografer_steering-by-example_2022}, 
(ii) produce sufficiently powerful classification models from relatively small inputs, allowing its use early on during the progression~\cite{kotsiantis2013decision}, and 
(iii) can be easily translated into steering rules, which represent the features of the relevant data~\cite{dimitriadou2014explore}.

Once the decision tree has been trained on the vulnerability features, the goal of the \emph{vulnerability features extraction} phase is to retrieve the rules to steer the next StatAG generation from the structure of the trained decision tree.
We name them \emph{steering rules} and must consider the values of the vulnerability features that generated only the ``relevant'' paths class.
To do so, we consider that the decision tree is built by using the vulnerability features as split points, and a leaf of the tree identifies whether or not the vulnerability features from the root to the leaf generate ``relevant'' paths.
This makes it possible to compute the inverse mapping from the ``relevant'' leaves to the root, keeping track of the vulnerability features values that only they have in common.
We can build the set of steering rules considering that each path from the root to a ``relevant'' leaf node represents a \emph{conjunction} of decision rules that must be satisfied for the decision tree to classify it as relevant.
Consequently, the logical \emph{disjunction} of all such decision rules generates the set of steering rules. 

In the next step of the approach, we use the steering rules to consider only the vulnerabilities that match them before sampling the graph again with random walks, as the StatAG generator mandates.
In this way, the newly generated attack paths have a high probability of matching the ``relevant'' class for the attack path query given in input, thus accelerating the progression of the AG generation towards attack paths answering the query.

Let us note that while the system is running, it continues to use the latest retrieved steering rules until there is a slowdown in the precision of the generation process, indicating that the relevant attack paths have been exhausted. To check that this effectively means having reached the exact answer to the query, another training is launched to adjust the steering with new rules that take into account the cumulative collection of attack paths, effectively repeating the described process until it converges to the exact result (i.e., a new training does not return any additional attack paths).
At each iteration of SteerAG, preliminary attack path analysis can be performed to answer the query, until all paths are generated. The process is even faster in convergence, guaranteeing analysis results that are identical in quality to the ones conducted on the fully generated AG.
%

\subsection{SteerAG Validation}
For the validation of SteerAG, we use the setting configurations used for StatAG and described in Section~\ref{sec:setting_config}, with the addition of another experimental parameter that is a set of 100 attack path queries obtained with different combinations of attack path features in the query and varying their ranges from 0 to 1 with steps of 0.1. This resulted in a total number of 60,000 experiments.

The main evaluation metrics for the steering approach are the \emph{recall} and the \emph{precision}.
The former, measured as $\frac{relevant\_paths}{all\_paths}$, informs about the convergence rate of the partial AG to the GT, while the latter, evaluated as $\frac{relevant\_paths}{retrieved\_paths}$, informs about the ability of the decision tree model to correctly retrieve the attack paths that are relevant for the query at every iteration.

\begin{figure*}[!ht]
    \centering
    \subfloat[Recall.\label{fig:recall}\centering]{{\includegraphics[width=0.45\linewidth]{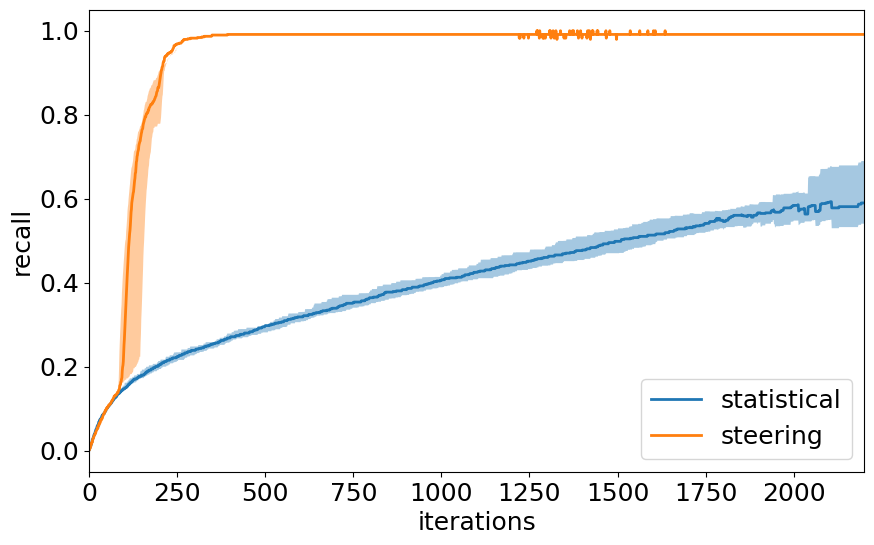}}}
    \qquad
    \subfloat[Precision.\label{fig:precision}\centering]{{\includegraphics[width=0.45\linewidth]{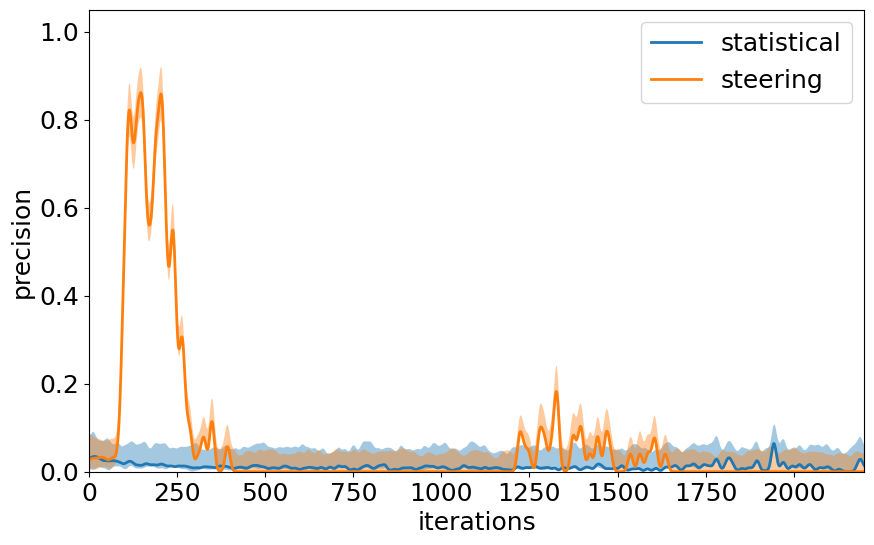}}}
    \qquad
    \caption{Comparison of (a) recall and (b) precision for StatAG and SteerAG. Median values are with full-color hue, and alpha blended areas indicate quartiles.}
    \label{fig:ml_metrics}
\end{figure*}

Fig.~\ref{fig:ml_metrics} reports the recall and precision of the performed experiments.
The median values trend for each curve is reported with full-color hue, while the alpha blended areas identify the variations between the upper and lower quartile values. 
%
From the recall trend (Fig.~\ref{fig:recall}) we can observe the very good performance of SteerAG that reaches values very close to 1 after just 250 iterations, with the activation of the steering mechanism starting at iteration 100 on average.
In particular, we can recognize two thresholds during SteerAG: between iteration 100 and 150 we observe early partial results (recall lower than 0.5 but fastly increasing), useful for exploration; from iteration 200 until iteration 250 (11\% of the execution) the results are mature (recall higher than 0.9), supporting early decision making; finally, 
from iteration 250 on we observe definitive partial results (recall higher than 0.98). 

It is interesting to note that without the steering approach, the convergence to the GT is very slow: at the definitive partial results threshold (iteration 250), random sampling has achieved a recall value of just 0.2.
Comparing the two trends, we can conclude that the steering approach accelerates the convergence to the GT, saving around 90\% of the iterations consistently in all the experiments tested, providing a more timely analysis of attack paths.

While the recall describes the ability of SteerAG to rapidly converge to the GT, 
the precision informs the analysts about the percentage of attack paths answering the query at each iteration: thus, high values correspond to many paths answering the query.
Fig.~\ref{fig:precision} shows the median precision trend during the framework execution.
We can observe that the precision values are consistently high after the activation of the steering mechanism at iteration 100, presenting median values that reach peaks of 0.85.
Similarly to the recall trend, precision presents rapid growth in the early iterations, particularly during the generation of early partial results; this shows that the analyst can expect a fast transition from early to mature partial results. 
At iteration 500 (23\% of the execution) it slows down.
After 700 iterations (at 55\% of the execution), there is another peak of precision up to 0.2 (visible also in the recall trend in Fig.~\ref{fig:recall}.
It is due to the retraining of the decision tree to adjust the steering rules after the detection of the precision breakdown.
Indeed, a precision breakdown happens only when most of the paths are already retrieved, aligning to low values typical of random sampling (end of steering phase). 
The fact that the precision has lower median values in the second steering activation than in the first one can be explained by a lower residual probability of finding the few relevant attack paths left (less than 2\% at that stage).
In contrast, StatAG without the steering mechanism has a constant trend of low precision values because, as expected, the probability of uniformly randomly picking relevant attack paths to the issued query is much lower. 
Finally, the performance is not significantly affected by the stringency of the query (i.e., different query ranges) as we experimentally proved in Appendix~\ref{appendix:query}.

\section{Case Study Evaluation}
\label{sec:case_study}


\subsection{Application to Large Real Networks}
\begin{wraptable}{r}{0.4\textwidth}
\vspace{-20pt}
\caption{SoA settings.}
\label{tab:valid_sota}
\resizebox{0.4\textwidth}{!}{%
\begin{tabular}{l|c|c|c|}
\cline{2-4}
 & \textbf{\begin{tabular}[c]{@{}c@{}}num. \\ hosts\end{tabular}} & \textbf{\begin{tabular}[c]{@{}c@{}}num.\\ vulns\end{tabular}} & \textbf{\begin{tabular}[c]{@{}c@{}}all\\ paths\end{tabular}} \\ \hline
\multicolumn{1}{|r|}{Wang et al. (2016)~\cite{wang_attack_2016}} & 7 & 2 & \checkmark \\ \hline
\multicolumn{1}{|r|}{Tian et al. (2017)~\cite{tian_network_2017}} & 4 & ND & \checkmark \\ \hline
\multicolumn{1}{|r|}{George et al. (2018)~\cite{george_graph-based_2018}} & 50 & ND & $\times$ \\ \hline
\multicolumn{1}{|r|}{Yichao et al. (2019)~\cite{yichao_improved_2019}} & 50 & 5 & $\times$ \\ \hline
\multicolumn{1}{|r|}{Li et al. (2019)~\cite{li_concurrency_2019}} & 18 & 7 & \checkmark \\ \hline
\multicolumn{1}{|r|}{Li et al. (2020)~\cite{li_implementing_2020}} & 15 & 12 & \checkmark \\ \hline
\multicolumn{1}{|r|}{Sun et al. (2022)~\cite{sun_heuristic_2022}} & 8 & 4 & $\times$ \\ \hline
\multicolumn{1}{|c|}{Proposed approach} & >250 & >50 & \checkmark \\ \hline
\end{tabular}%
}
\vspace{-20pt}
\end{wraptable}
The first application we prove deals with scalability.
In Table~\ref{tab:valid_sota} we report the works of the literature that address the generation and analysis of attack paths, highlighting the considered maximum number of hosts, vulnerability per host, and whether they generate all paths. 
In this last case, the approaches define heuristics to prioritize the paths under analysis and avoid the full generation (more details in Section~\ref{sec:related}).
From Table~\ref{tab:valid_sota} we can observe that the largest experimental settings include either many hosts (50) and few vulnerabilities (5), or a medium number of hosts (15) and vulnerabilities (12 vulnerabilities per host).
For the validation of our approach, we already considered networks bigger than those, but here we report an even bigger and real network.
We emulated the network of a university department composed of around $250$ devices arranged in $5$ LANs connected through routers.
The network has been scanned for vulnerabilities using commercially available vulnerability scanners (i.e., Tenable Nessus).

\begin{wrapfigure}{l}{0.5\textwidth}
    \vspace{-20pt}
    \centering
    \includegraphics[width=0.9\linewidth]{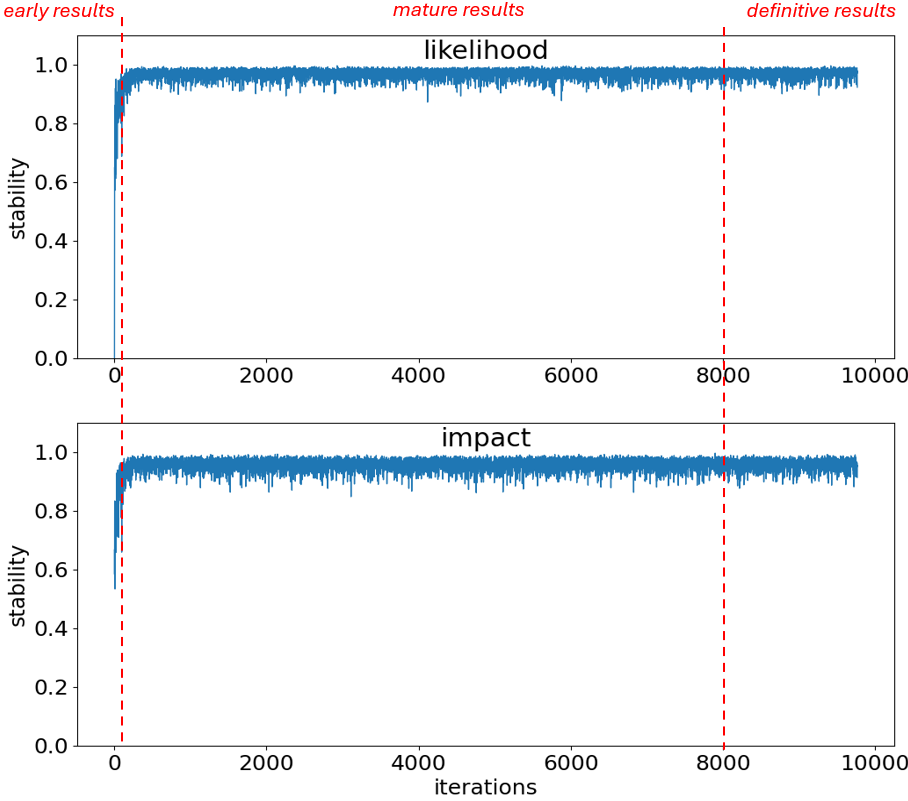}
    \caption{AP features stability trend.}
    \label{fig:real_stability}
    \vspace{-20pt}
\end{wrapfigure}
Concerning the attack path query and for the sake of example, we consider the most common risk analysis that requires attack paths with likelihood and impact higher than 0.9, as the most risky situation. We prove later the support of a wide range of analysis queries.
Let us note that the generation of the complete AG is intractable for such a size with the classic full generation, but we show that the progressive generation allows having controllable partial results.
%
The stability trend of Fig.~\ref{fig:real_stability} shows that until iteration 100 the results are early partial, with low and fluctuant stability indicating a still high approximation of the complete AG. 
From iteration 100, high stability values start both for likelihood and impact, thus hypothesizing that it is the time of significant partial results, and support tasks like hypotheses formulation.
However, it is reasonable to observe a constant trend of the stability values before considering the results statistically significant:
a reasonable time to consider mature partial results is at iteration 500, where both likelihood and impact show a regular median stability of 0.85, progressively increasing. This phase supports hypothesis testing, comparative or confirmatory analyses, and early decision-making.
From this time, the attack path analysis can start, considering that it approximates at least 85\% of the complete AG, until reaching very stable results (definitive partial) at iteration 8000.

\begin{wrapfigure}{r}{0.5\textwidth}
    \vspace{-20pt}
    \centering
    \includegraphics[width=\linewidth]{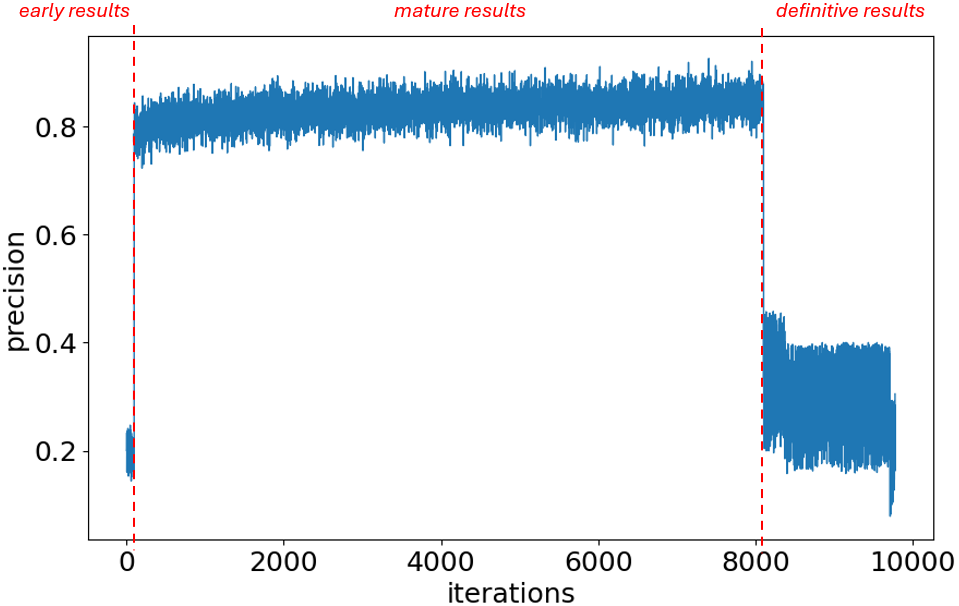}
    \caption{Precision trend SteerAG.}
    \label{fig:real_precision}
    \vspace{-20pt}
\end{wrapfigure}
Meanwhile, the analyst observes the precision trend of Fig.~\ref{fig:real_precision}.
She notices a rapid growth at iteration 100, indicating that the SteerAG has been activated, thus indicating the starting point of mature partial results.
The precision has an average value of 0.8 until around iteration 8000.
During this period, the analyst can perform attack path analysis, considering that new paths are incoming to answer the query.
When the precision breaks down around iteration 8000, the results are very close to the exact ones (definitive partial results).
These observations agree with the stability trend (Fig.~\ref{fig:real_stability}).
%
To give an idea of the advantages provided by this framework in terms of time, an analyst uses the progressive generation to perform early decision-making in interactive times. 
Each iteration has an average duration between 0.5 and 5 seconds, and after a few minutes ($\approx 5$) from the beginning, she can start the analysis with mature partial results for any issued query.

\subsection{Coverage of Attack Path Analyses}
\begin{wrapfigure}{l}{0.47\textwidth}
    \vspace{-20pt}
    \centering
    \includegraphics[width=\linewidth]{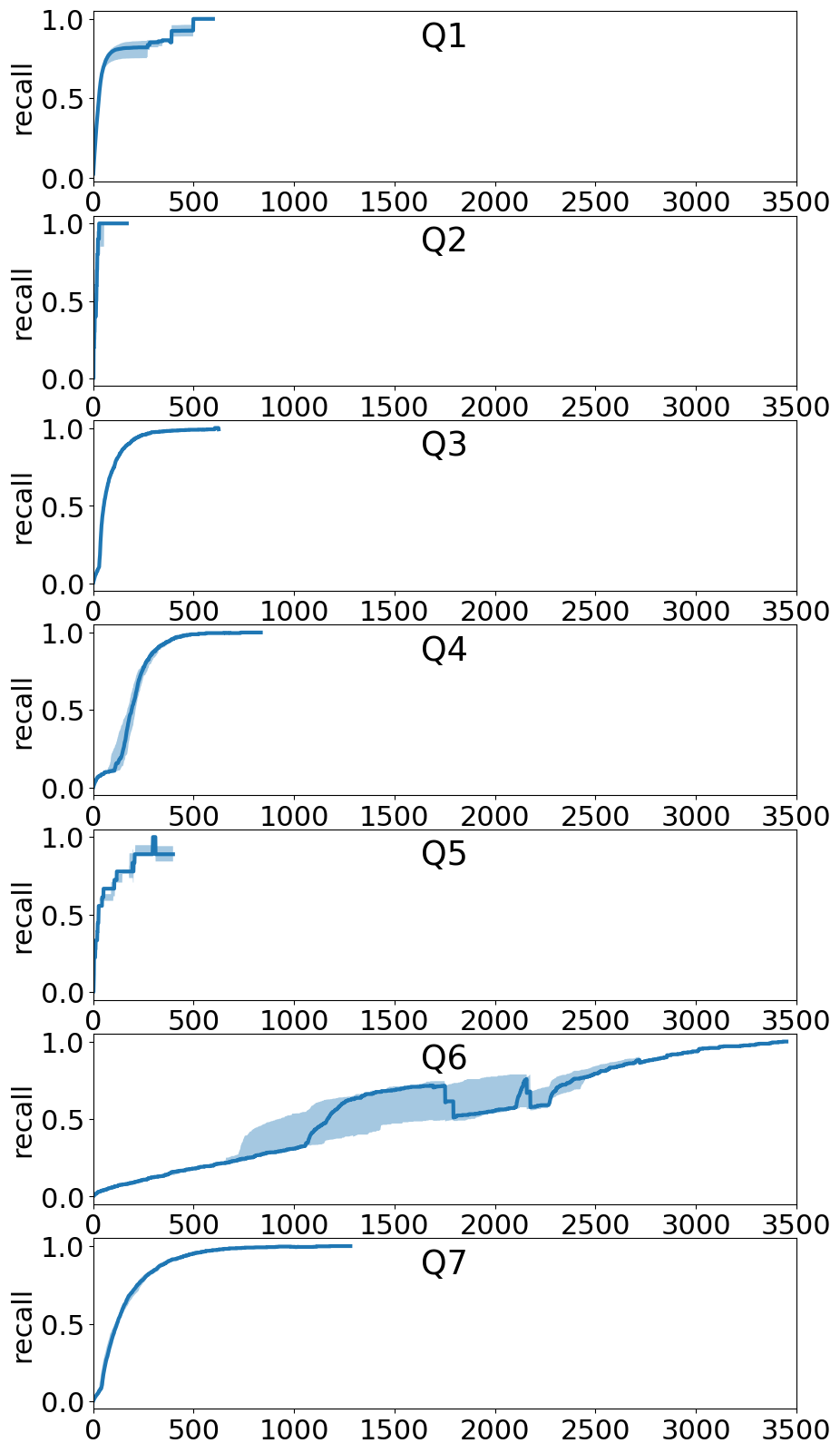}
    \caption{Recall for different queries.} \label{fig:sok_queries}
    \vspace{-20pt}
\end{wrapfigure}

To conclude the case study, we analyze how the proposed approach can analyze a selected set of queries with the configuration settings used for the validation (Section~\ref{sec:setting_config}).
The attack path queries are extracted from state-of-the-art works addressing the systematization of attack path analysis~\cite{landoll2017information,1004377}:\\
\textbf{Q1}: Evaluate the risk on the shortest attack paths~\cite{noel2014metrics};\\
\textbf{Q2}: Retrieve the attack paths with maximum impact (i.e., impact~=~1)~\cite{linawati_security_2014};\\
\textbf{Q3}: Retrieve the attack paths with maximum likelihood (i.e., likelihood~=~1)~\cite{kavallieratos_attack_2020}\\
\textbf{Q4}: Retrieve the attack paths with maximum risk (i.e., likelihood~$\cdot$~impact~=~1)~\cite{9519490};\\
\textbf{Q5}: Retrieve the attack paths corresponding to \emph{black swan} attacks, i.e., those ones with high impacts (impact~$>$~0.9), but unlikely to happen (likelihood~$<$~0.3)~\cite{aven2013meaning};\\
\textbf{Q6}: Retrieve the attack paths corresponding to \emph{gray swan} attacks, i.e., those with very low impacts (impact~$<$~0.3), but highly probable (likelihood~$>$~0.9)~\cite{khakzad2015major};\\
\textbf{Q7}: Prioritize the attack paths by risk~\cite{zenitani_attack_2023}. Let us note that it corresponds to an ensemble of queries, retrieving first the paths with risk 1, then between 0.9 and 1, and so on. To avoid the complete enumeration of the paths, we stop the retrieval at risk 0.5.
Fig.~\ref{fig:sok_queries} reports the recall median trend for the analysis of these queries, where each experiment is repeated 100 times. 

%
%
\noindent The recall trends in Fig.~\ref{fig:sok_queries} show that most queries (specifically, Q1-Q5 and Q7) converge very rapidly to the GT, indicating the concrete support the approach can provide during attack path analysis.
Among them, Q2, Q3, and Q4 ask for precise values of attack path features, while Q1, Q5, and Q7 ask for more complex queries, requiring more than one traversal in the same AG.
In particular, query Q6 for gray swan attacks has the worst performance, requiring 3500 median iterations to retrieve all the paths.
The reason for its worse performance is the need to perform more traversals and the many paths answering the query.
Let us remark it still provides approximate answers but with slower convergence.
This case study showed the capabilities of the approach to perform attack path analysis with networks' size that existing solutions cannot address and according to state-of-the-art attack path analysis.
As a final consideration, let us remark that the queries can be issued anytime during the execution and the approach triggers the steering mechanism upon the arrival of a query.

\section{Related Work}
\label{sec:related}

\noindent {\bf Attack Graph Generation.}
To address the scalability issue of AGs, one possible solution consists of leveraging distributed and parallel computing.
The distributed approaches, as Kaynar et al.~\cite{kaynar_distributed_2016} and Sabur et al.~\cite{sabur_toward_2022}, partition the network based on its services and vulnerabilities and assign the different partitions among distributed agents.
Each agent then computes the attack paths on smaller portions of the AG which are finally combined.
The approaches based on parallel computing~\cite{li_concurrency_2019,li_implementing_2020} adapt the serial graph search algorithms to multi-core environments. 
Our approach is agnostic to distribution or parallelization, as it acts on the AG workflow.

Another possibility is using Artificial Intelligence (AI).
For example, Li et al.~\cite{li_deepag_2022} leverage Deep Learning and node2vec~\cite{grover_node2vec_2016} to train a neural network with data from system logs and use it to predict dependencies between network hosts and label the sequence of events as potential attack paths.
Besides, Ychao et al.~\cite{yichao_improved_2019} model the path discovery problem as a planning problem on graphs.
They translate vulnerabilities into formal actions and use state-of-the-art planners to discover attack paths.

While Distributed and AI-based solutions provide a good improvement in the scalability of AG generation, they still require waiting until the generation is completed to perform the first analyses.
In dynamic scenarios where network components change rapidly (and so do attack graphs), it is not reasonable to wait for the re-computation of the attack paths each time the environment changes.
In contrast, the proposed approach allows for querying Attack Path analyses at any time with quality-controlled statistically significant results.\\

\noindent {\bf Attack Path Analysis.}
Most current AG-based systems have tackled the scalability issue in cyber risk analysis by prioritizing specific attack paths based on security needs.
Some of them prioritize the analysis considering the graph's topological structure, as Sun et al.~\cite{sun_heuristic_2022} who use nodes' degree, and Gonda et al.~\cite{gonda2018analysis} who map AGs to planning graphs and leverage the nodes' centrality.
Other works leverage security metrics to map potential exploits to the easiest reachable attack target~\cite{liu_goal-oriented_2010} or to prioritize attack paths on the security conditions~\cite{feng_generation_2017} and their cost values~\cite{feng_generation_2017}.
%
These works take specific assumptions on priorities and, therefore, are fine only for scenarios that conform to them.
Moreover, they are static and do not follow the network exposure.

Few works addressed the problem of considering attack path analysis to prioritize the generation.
Yuan et al.~\cite{yuan_attack_2020} and Salayma et al.~\cite{salayma_threat_2023} model the AG through a graph database and perform attack path analysis by suitably writing queries to the database.
While these solutions allow the expression of complex customizable queries, they have the drawback that they depend on the graph database, which may slow down the performance during the analysis~\cite{guia2017graph}.
Differently, Nadeem et al.~\cite{nadeem_alert-driven_2022} use intrusion alerts to drive the attack path analysis.
They translate alert events to episode sequences used to build the AG and execute the analysis driven by alerts.
Similarly, Hassan et al.~\cite{hassan2020tactical} perform rule matching on system logs to identify the events that match attack behaviors described in security knowledge bases.
%
While these alert-based approaches are context-aware, they may not represent the entire environment, such as network components with no alerts detected.
In contrast, the proposed approach assesses the entire network without introducing any bias to a fixed analysis. On the contrary, it dynamically adapts to different queries.
\section{Discussion and Concluding Remarks}
This paper proposed a new framework to generate and analyze AGs based on progressive data analysis.
We developed StatAG to generate statistically significant partial AGs and SteerAG to accelerate AG generation based on attack path queries. 
Our approach adapts better to network changes without recomputing the entire AG, a feature lacking in traditional AG generation methods. 
Extensive experiments and a case study validate its capability to analyze large networks and efficiently support common attack path analyses.
While our approach is complementary to existing methods that can be integrated without loss of generality, to the best of the authors' knowledge, it is the first contribution to advancing the classic AG-based process.
All the materials and the source code of all components are publicly available\footnote{\url{https://github.com/XAIber-lab/ProgressiveAttackGraph}}.
%

Some implications of this work are worth discussing.
Although our approach is applied to \emph{topological} AGs, we believe it can be easily adjusted for attack path analysis over logical ones, for example, adapting the path construction with model-checking~\cite{clarke1997model}.
%
Additionally, we considered the likelihood and impact as attack path features since they are the main ones used in the literature on cyber risk management. Still, others could be considered, dependent on the application domain (e.g., cloud~\cite{pauley2022measuring} and automotive~\cite{macher2016review}).
We also suggest exploring other ML models, e.g., Graph Neural Networks~\cite{9046288} that, while allowing to capture non-linear relations, are challenging to adapt to our approach due to their resource-intensive training and their black-box nature that makes it difficult to retrieve the steering rules.
%

Looking at the research opportunities the proposed work opens, one concerns the management of progressive thresholds to indicate early, mature, and definitive partial results.
Future works may study automated ways to set progressive thresholds based on real-time indicators. Another opportunity involves the improvement of the decision tree predictions after the precision breakdown. Currently, we re-train the decision tree with the newly incoming paths, but more advanced strategies to optimize the re-train of the ML model may be adopted. 
%
Finally, while the proposed framework can be included in an automatic pipeline, where a process issues analysis queries, it opens the research for the design of human-in-the-loop systems.

\subsubsection*{Acknowledgements.}
This work was partially supported by project SERICS (PE00000014) under the MUR National Recovery and Resilience Plan funded by the European Union - NextGenerationEU.
%
%
%
%
\bibliographystyle{splncs04}
\bibliography{samplepaper}
%





\appendix
\section{Query stringency analysis}
\label{appendix:query}
\begin{figure}[!ht]
    \centering
    \subfloat[Recall trend.\label{fig:query_recall}\centering]{{\includegraphics[width=0.28\linewidth]{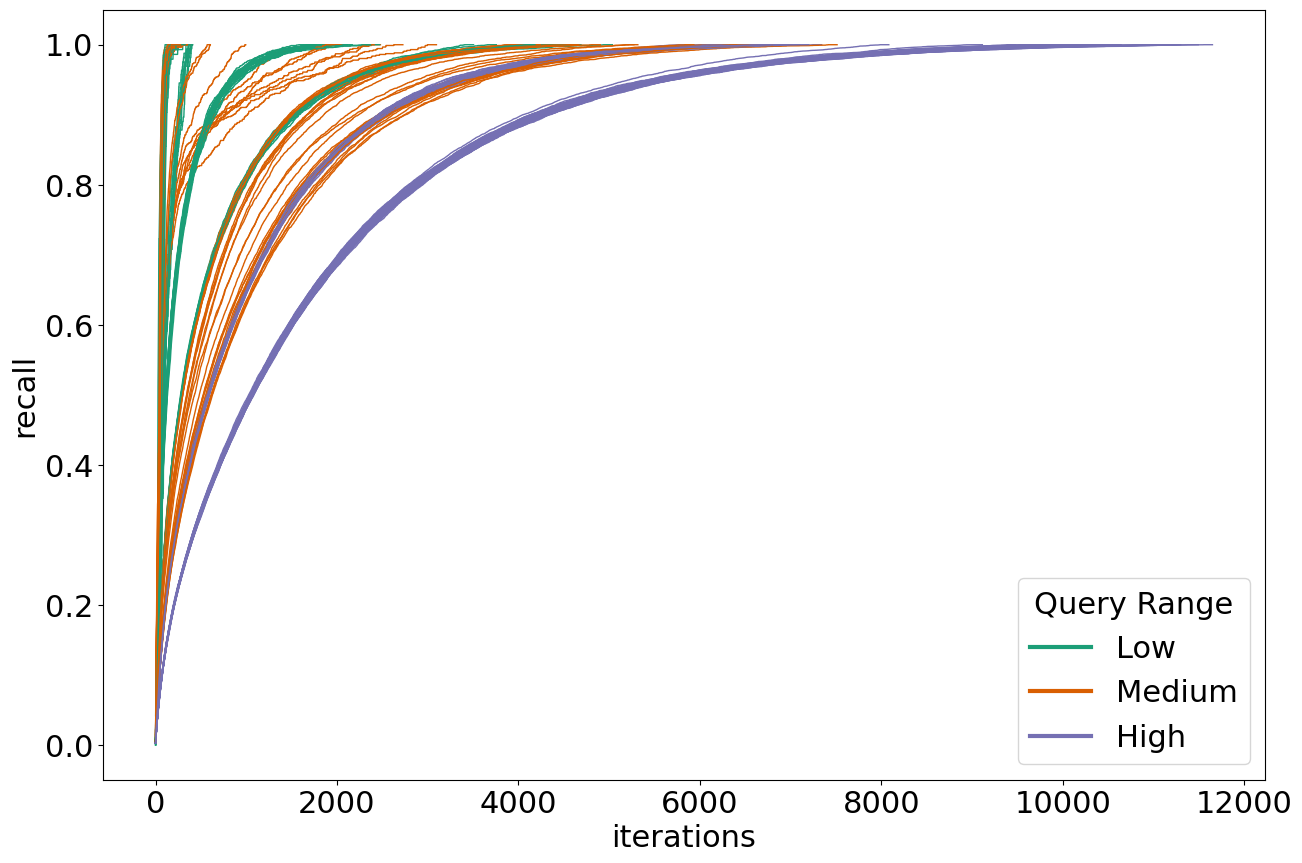}}}
    \qquad
    \subfloat[Convergence speed.\label{fig:query_convergencerate}\centering]{{\includegraphics[width=0.28\linewidth]{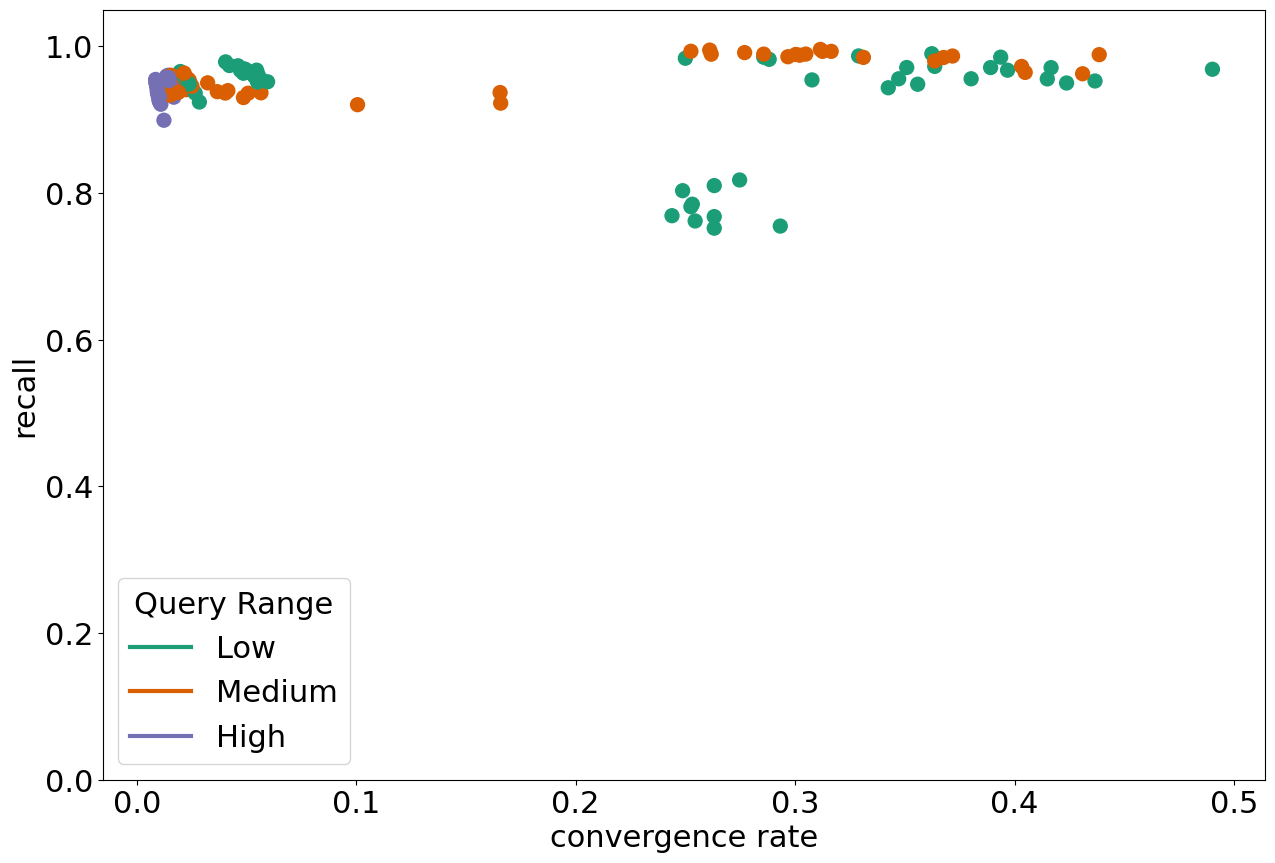}}}
    \qquad
    \subfloat[Missing rate.\label{fig:query_distance}\centering]{{\includegraphics[width=0.28\linewidth]{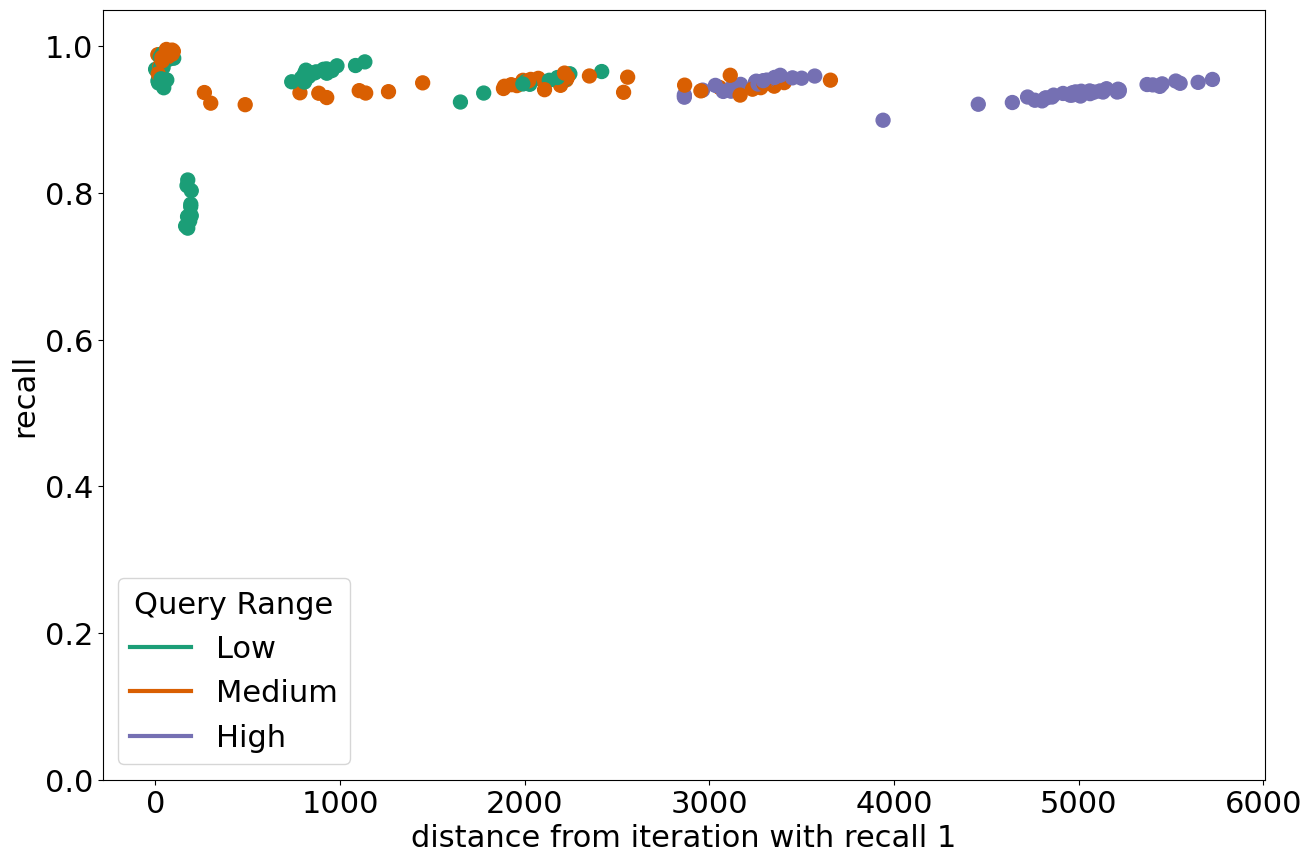}}}
    \qquad
    \caption{Analysis of steering mechanism for different query ranges.}
    \label{fig:query_analysis}
\end{figure}

We are interested in studying how a stricter query affects performance to avoid situations where a very specific query gets stuck due to slow convergence.
To validate this, we define three ranges of queries: 
\emph{low}, when the values of the attack path features are in a range of at most 0.2 (strict query);
\emph{medium}, when it is between 0.3 and 0.5;
and \emph{high}, when it is over 0.5 (large query).
We report different analyses in Fig.~\ref{fig:query_analysis}.
\vspace{-20pt}
\vspace{-20pt}
Fig.~\ref{fig:query_recall} shows the recall trend of the different experiments for each query range.
It highlights three different trends that for the low, medium, and high ranges are respectively the fastest, medium, and slowest convergence to the GT.
To quantify the convergence speed to the GT, in Fig.~\ref{fig:query_convergencerate} we plot the recall values at the point of the maximum convergence of the recall curves. In practice, it corresponds to the peak of the recall curve before its stability.
If the maximum convergence rate is low ($< 0.2$) for high recall values ($> 0.8$), then it means that the approach converges slowly to the GT. It is the case of 100\% of the experiments with high query ranges, 54\% of the experiments with medium ranges, and 25\% of the ones with low query ranges.
%
In contrast, the rest of the medium and low-range queries show high convergence speed ($>0.3$) when the recall is higher than 0.8, indicating quick convergence to the GT. 
%
In Fig.~\ref{fig:query_distance}, we analyze the iterations remaining from peak convergence to retrieve all relevant paths (missing rate). The left part of the plot comprises experiments closer to recall value $1$, including all low-range experiments and 70\% of medium-range ones. The right one involves cases needing more iterations to reach recall 1, encompassing all high-range experiments and the remaining medium-range ones.
%
Results for low and medium ranges confirm SteerAG's ability to quickly retrieve accurate attack paths even for strict queries. High-range query results seem inferior, but this is due to the large number of attack paths to retrieve, influenced by the sampling size. Precision remains consistently high, reaching convergence rates with recall values exceeding 0.8, even for high-range queries.
Overall, we do not observe an effect on the performance. 


\end{document}